# ESTIMATION BAYESIENNE D'UN MODELE HYDRODYNAMIQUE 1D D'UNE RIVIERE INFLUENCEE PAR LA MAREE : APPLICATION A LA SEINE AVAL, FRANCE

*Bayesian estimation of a 1D hydrodynamic model in a tidal river: Application to the Lower Seine River, France*

**Auteurs :** MENDEZ RIOS Felipe[1], LE COZ Jérôme[2], RENARD Benjamin[3], TERRAZ Théophile[4]

[1] INRAE, UR RiverLy, centre Lyon-Grenoble-Auvergne-Rhône-Alpes, 5 rue de la Doua CS 20244, 69625 Villeurbanne, France, e-mail: felipe-alberto.mendez-rios@inrae.fr
[2] INRAE, UR RiverLy, centre Lyon-Grenoble-Auvergne-Rhône-Alpes, 5 rue de la Doua CS 20244, 69625 Villeurbanne, France, e-mail: jerome.lecoz@inrae.fr
[3] INRAE, UMR RECOVER, centre Provence-Alpes-Côte d'Azur, 3275 Route de Cézanne, CS40061, 13182 Aix en Provence, France, e-mail: benjamin.renard@inrae.fr
[4] INRAE, UR RiverLy, centre Lyon-Grenoble-Auvergne-Rhône-Alpes, 5 rue de la Doua CS 20244, 69625 Villeurbanne, France, e-mail: theophile.terraz@inrae.fr

**Choix du thème/session :**
2/ Prévision des crues et des inondations

**Mots clefs**:
Calage, modèle hydrodynamique 1D, niveau d'eau, débit, incertitude, coefficient de rugosité, marée.

## 1. RÉSUMÉ

La surveillance du niveau d'eau ainsi que du débit au niveau des stations hydrométriques est essentielle pour la détection et la prévision des crues et des inondations. La mesure en continu est possible pour le niveau d'eau, alors que le débit doit être calculé et non mesuré directement. C'est pourquoi plusieurs méthodes ont été développées comme la mesure de vitesse en surface [1] ou les courbes de tarages [2][3]. Néanmoins, les stations hydrométriques peuvent être influencées par la marée, ce qui provoque un remous variable en régime transitoire. En régime quasi-permanent, le remous variable est gérable par une relation hauteur-dénivelée-débit [4] basée sur la formule de Manning-Strickler et des mesures du niveau d'eau et de la pente de la ligne d'eau. Mais en régime fortement transitoire lorsque l'effet de la marée est prononcé, ce type de relation ainsi que des variantes se sont avérées peu performantes [5]. Pour modéliser la dynamique complexe de l'écoulement, y compris lors de l'inversion du flux, une approche via l'estimation bayésienne d'un modèle hydrodynamique 1D est proposée. Ici, l'estimation d'un modèle s'entend comme l'estimation de la distribution a posteriori des paramètres et du modèle d'erreur structurelle.

Afin d'établir le modèle hydrodynamique 1D, la géométrie des sections en travers, le coefficient de résistance hydraulique par tronçon, le ou les débits en amont et le niveau d'eau en aval sont nécessaires. Dans la modélisation hydrodynamique, la résistance hydraulique (via un jeu de coefficients de Strickler) est le principal paramètre à caler [6], mais le calage manuel est rendu difficile en raison des variations spatiales de la rugosité et de la nature transitoire de l'écoulement. En outre, la compréhension et la quantification des incertitudes associées aux données et au modèle constituent une étape importante du processus de calage. Par conséquent, un calage automatique des coefficients de rugosité est proposé via l'inférence bayésienne.

En termes d'outils numériques, le code hydrodynamique 1D utilisé ici est Mage [7], développé par INRAE, qui résout les équations 1D de Saint-Venant pour les écoulements fluviaux et transitoires. Toutefois, la méthode proposée n'est pas spécifique d'un code de simulation donné : elle peut s'appliquer à tout code hydrodynamique 1D usuel. Le calage bayésien se réalise via le logiciel *BaM!* [8] (*Bayesian modeling* : https://github.com/BaM-tools) qui permet de spécifier des informations a priori sur les paramètres d'un modèle (ici les coefficients de résistance hydraulique) pour ensuite les estimer avec leur incertitude associée, en utilisant des observations elles-mêmes munies de leur incertitude (ici non seulement des niveaux d'eau, mais aussi les campagnes de jaugeages).

## 2. APPLICATION

La Seine aval en France est prise comme cas d'application, car il s'agit d'un modèle hydraulique simple (un unique tronçon sans débordement ni casier hydraulique) avec un fort effet de marée, et des campagnes de jaugeage sur plusieurs cycles de marée et des limnigrammes disponibles [9] [10]. Quinze marégraphes sont présents sur le domaine de simulation du modèle d'une longueur totale de 139 kilomètres entre la station de Poses (amont) et l'aval de celle de Tancarville [11].

Un seul coefficient de résistance hydraulique est défini sur le bief en amont de la confluence avec l'Eure alors que différents coefficients constants sur 11 tronçons sont définis en aval de la confluence. Les séries temporelles de débit de la Seine à Poses et de l'Eure, le seul affluent significatif, sont spécifiées comme conditions aux limites amont. La condition limite aval est la série temporelle des hauteurs d'eau de la Seine à l'aval de Tancarville, reflétant le signal de la marée. Le calage est effectué à partir des observations du débit provenant de la campagne de jaugeages ADCP faite par le Service de Prévision des Crues (SPC) DREAL Normandie en septembre 2015 à Rouen, pendant un cycle de marée. Pour la validation, les données des marégraphes et des jaugeages à Rouen, Heurteauville et Aizier ont été sélectionnées.

La distribution a priori de chaque coefficient de Strickler par tronçon est définie comme log-normale avec un intervalle de probabilité à 95% égal à [33 ; 49] (paramètres ln (40.5) et 0.1), pour couvrir les valeurs typiquement attendues pour ce type de cours d'eau. L'erreur structurelle du modèle est supposée indépendante d'une observation (jaugeage) à l'autre, gaussienne, de moyenne nulle et d'écart-type inconnu. Cet écart-type sera estimé en même temps que les coefficients de Strickler.

L'estimation bayésienne fournit alors les distributions a posteriori des 12 coefficients de Strickler du modèle, représentées par un grand nombre d'échantillons (10 000) générés via un algorithme de Monte Carlo par Chaînes de Markov (MCMC). Ces échantillons sont utilisés pour non seulement identifier les coefficients "maxpost" (maximisant la densité a posteriori) considérés comme optimaux, mais aussi quantifier et propager leur incertitude. Par la suite, une propagation s'effectue afin d'estimer les séries temporelles du niveau d'eau et du débit de toutes les sections en travers du modèle avec l'incertitude associée.

## 3. RÉSULTATS

L'étape de calage permet d'estimer la distribution de probabilité jointe, ainsi que le vecteur de paramètres optimal. Il faut remarquer que lors le cadre de l'inférence bayésienne, le but est plutôt d'identifier la distribution au lieu d'une valeur optimale. Dans la Figure 1, on constate que certaines distributions a posteriori restent très similaires à l'a priori, notamment pour les tronçons 1, 6, 7 et 10, ce qui signifie que les observations ne fournissent pas plus d'information que les connaissances a priori. En revanche, le tronçon bénéficiant des observations pour le calage est généralement mieux identifiable. Cette situation peut également se produire pour d'autres tronçons, même en l'absence de donnés d'intervention lors du calage.

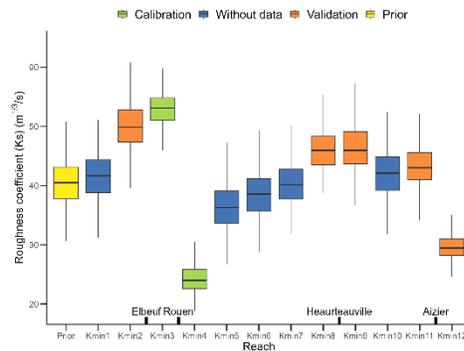

*Figure 1. Boîtes à moustaches des distributions a priori (en jaune) et a posteriori (en vert si le tronçon ayant bénéficié d'observations pour le calage, et en orange pour les stations utilisées pour la validation) des coefficients de Strickler des 12 tronçons du modèle hydrodynamique 1D de la Seine aval.*

L'étape de prédiction consiste à utiliser le modèle après calage pour réaliser des calculs de hauteur et de débit avec leurs incertitudes sur toutes les sections transversales définies dans le modèle. La Figure 2 permet de comparer les simulations aux observations utilisées pour le calage avec l'incertitude quantifiée via QRevInt par la méthode Oursin [12] à un intervalle de confiance de 95%.

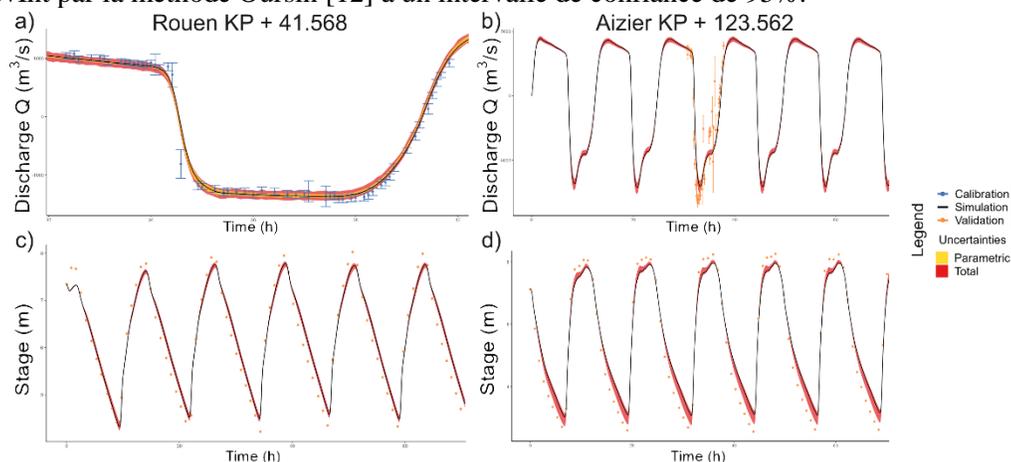

*Figure 2 : Résultats du calage du modèle 1D de la Seine aval : a) Hydrogramme à Rouen et b) Aizier et limnigrammes à b) Rouen et d) Aizier simulés avec leurs incertitudes à 95%*

L'incertitude des débits, exprimée comme l'écart-type moyen de la distribution sous-jacente (quelle que soit cette distribution), est de l'ordre de 37 m$^3$/s à la station de Rouen, pour une amplitude totale de variation des débits d'environ -1000 à 1000 m$^3$/s. A la station d'Aizier, cette incertitude est de l'ordre de 110 m$^3$/s pour des variations comprises entre -8000 m$^3$/s à 5000. En termes de niveau d'eau, ce même écart-type varie d'environ 0.03 m à Elbeuf jusqu'à 0.07 m à Aizier. Les observations de débit avec leurs incertitudes ont tendance à rester à l'intérieur des enveloppes d'incertitude de la simulation, ce qui suggère une bonne performance des résultats de la modélisation pour enregistrer la dynamique de la marée, notamment à la station utilisée pour le calage. Néanmoins, cette performance est affaiblie lorsqu'il s'agit de valider les résultats sur d'autres stations.

## 4. CONCLUSIONS

La méthode de calage bayésienne se présente comme une solution avec un potentiel intéressant pour le calage automatique d'un modèle hydrodynamique 1D en écoulement transitoire prenant comme données de calage des mesures de niveau d'eau et de débit avec leurs incertitudes. De plus, cette approche permet de réaliser des prédictions avec incertitudes (dues au calage) quantifiées, notamment en contexte de prévision puisque l'information dont le modèle a besoin (débits amont, niveau aval) peut être prévue. Le cas d'application de la méthode a été les stations influencées par la marée, toutefois la méthode pourra être étendue à d'autres secteurs, tout tronçon surveillé en prévision des crues, et même tout secteur modélisé en 1D. En conclusion, cette étude a pour but de contribuer à l'amélioration de la prévision des inondations et des systèmes d'alerte, et à la prise en compte des incertitudes associées.

## 5. PERSPECTIVES

Nous prévoyons d'étudier la performance de la méthode en testant d'autres configurations de calage, avec d'autres campagnes de jaugeages ADCP à d'autres stations hydrométriques, notamment réaliser le calage en utilisant seulement des mesures du niveau d'eau pour mettre en évidence l'importance des jaugeages pour le calage du modèle en hauteur et en débit. Enfin, le temps de calcul de l'étape de calage reste encore à optimiser, notamment en simplifiant le modèle utilisé à cette étape : à ce stade, le calage prend de l'ordre de deux jour; par contre, la prédiction de séries pendant la période d'analyse (d'environ deux jours) ne prend que quelques heures. Un plan d'expérience numérique est envisagé pour évaluer et optimiser la performance de la méthode, en vue de son application opérationnelle.

## 7. REMERCIEMENTS